\documentclass[twoside,twocolumn,english,aps]{revtex4}
\usepackage[T1]{fontenc}
\usepackage[latin9]{inputenc}
\usepackage[a4paper]{geometry}
\usepackage{amsmath}
\usepackage{amssymb}
\usepackage{graphicx}
\usepackage{esint}

\makeatletter
\@ifundefined{textcolor}{}
{%
 \definecolor{BLACK}{gray}{0}
 \definecolor{WHITE}{gray}{1}
 \definecolor{RED}{rgb}{1,0,0}
 \definecolor{GREEN}{rgb}{0,1,0}
 \definecolor{BLUE}{rgb}{0,0,1}
 \definecolor{CYAN}{cmyk}{1,0,0,0}
 \definecolor{MAGENTA}{cmyk}{0,1,0,0}
 \definecolor{YELLOW}{cmyk}{0,0,1,0}
 }

\makeatother

\usepackage{babel}
\begin{document}

\title{Universality and the three-body parameter of helium-4 trimers}

\author{Pascal Naidon$\,^{1}$, Emiko Hiyama$\,^{1}$, and Masahito Ueda$\,^{2}$}

\affiliation{$\,^{1}$RIKEN Nishina Centre, RIKEN, Wako 351-0198, Japan, }

\affiliation{$\,^{2}$Department of Physics, University of Tokyo, 7-3-1 Hongo,
Bunkyo-ku, Tokyo 113-0033, Japan}

\date{\today}
\begin{abstract}
We consider a system of three helium-4 atoms, which is so far the
simplest realistic three-body system exhibiting the Efimov effect,
in order to analyse deviations from the universal Efimov three-body
spectrum. We first calculate the bound states using a realistic two-body
potential, and then analyse how they can be reproduced by simple effective
models beyond Efimov's universal theory. We find that the non-universal
variations of the first two states can be well reproduced by models
parametrized with only three quantities: the scattering length and
effective range of the original potential, and the strength of a small
three-body force. Furthermore, the three-body parameter which fixes
the origin of the infinite set of three-body levels is found to be
consistent with recent experimental observations in other atomic species.
\end{abstract}
\maketitle

\section{Introduction}

The universal attraction found by V. Efimov \cite{rf:efimov} for
any quantum system of three particles interacting through short-range
interactions with a large scattering length has now been evidenced
in many experiments using ultra-cold atoms~\cite{rf:ferlaino,rf:kraemer,rf:barontini,rf:pollack,rf:huckans,rf:lompe1,rf:lompe2,rf:nakajima,rf:nakajima2,rf:ottenstein,rf:wenz,rf:williams,rf:zaccanti,rf:gross,rf:gross2,rf:Berninger}.
In particular, Efimov trimers, i.e. three-body bound states resulting
from this attraction, were observed as a function of the scattering
length. Because these trimers are unnaturally large compared with
the range of the interactions, they have universal properties determined
solely by the universal attraction and a few parameters. In particular,
their spectrum has a simple structure with discrete scale invariance.
This was confirmed experimentally to some degree, but there appeared
quantitative deviations from this structure.

One of the main reasons is the fact that at most the first two states
of the spectrum could be observed so far. It is known that these first
states do not follow accurately the universal behaviour expected for
the higher excited states (more loosely bound states), because their
size is not very large compared to the range of the interactions and
therefore they still depend on the details of these interactions.
However, it is quite involved to correctly describe these interactions
for three atoms because of their complex hyperfine structure and the
lack of knowledge of the three-body potential surfaces. For this reason,
experimental results have been interpreted so far using either interaction-dependent
corrections to the universal theory of Efimov~\cite{rf:platter,rf:Ji,rf:nakajima,rf:nakajima2},
or by other effective models reproducing the two-body physics in the
energy range of the observed trimers~\cite{rf:lee,rf:jonalasinio,rf:naidon}.

While these effective approaches could reproduce the experimental
results to some extent, it remains unclear on a theoretical basis
why they could do so. For example, corrections to the Efimov universal
theory sometimes required to introduce an ad-hoc variation of a 3-body
parameter to explain the data \cite{rf:nakajima,rf:nakajima2,rf:naidon}.
Another puzzling fact is that effective two-body model approaches
could reproduce some of the deviations from universal theory observed
in the experimental data, suggesting that they could be explained
by two-body interactions only \cite{rf:lee,rf:jonalasinio,rf:naidon}.
 While it is established that in general the knowledge of two-body
interactions only is not enough to accurately determine the energy
of Efimov trimers \cite{rf:dincao}, the contribution from realistic
two-body interactions to the short-range 3-body phase and non-universal
deviations are not fully understood.

The purpose of this paper is to clarify these issues by testing the
effective approaches with respect to the numerically exact solution
of a realistic theoretical model. We choose $\,^{4}\mbox{He}_{3}$,
as it is the simplest triatomic system with van der Waals inteactions
which exhibits the Efimov attraction~\cite{rf:Cornelius,rf:esrylin,rf:NielsenHelium,rf:Gonzalez.,rf:BraatenHelium}.

The paper is organised as follows. In section II, we review some of
the effective models used to describe Efimov trimer experiments. In
section III, we present realistic calculations for $\,^{4}\mbox{He}_{3}$,
and how they are reproduced by these effective models.

\section{Effective models for Efimov physics}

\subsection{Efimov's universal theory}

The essence of the Efimov effect is the appearance of an effective
attractive potential $-s_{0}^{2}/R^{2}$ attraction between three
particles with very large scattering length. Here, $s_{0}$ is a number
approximately equal to 1.00624 for identical bosonic particles, and
$R$ denotes the hyper-radius (average distance between particles),
\begin{equation}
R^{2}=\frac{1}{3}(r_{12}^{2}+r_{23}^{2}+r_{31}^{2}),\label{eq:Hyperradius}
\end{equation}
where $r_{12,}r_{23},\mbox{ and }r_{31}$ are the three particles'
relative distances. This attraction can lead to the existence of three-body
bound states, the so-called Efimov trimers. Because it is a long-range
attraction, trimers with sufficiently small binding energy extend
to large distances where the interactions are negligible. The only
effect of interactions is to set short-distance boundary conditions
on the three-body wave function. The first boundary condition occurs
when two particles come close to one another, but within a distance
$r$ larger than the range of their interaction. There, the wave function
$\psi$ has to satisfy the Bethe-Peierls boundary condition
\begin{equation}
\psi\;\underset{r_{ij}\to0}{\propto}\;\frac{1}{r_{ij}}-\frac{1}{a},\label{eq:BethePeierls}
\end{equation}
where $a$ is the s-wave scattering length of the two-body interaction,
which fixes the phase of the two-body wave function accumulated from
short distance at low energy. The second boundary condition occurs
when three particles come close together, but still at distances $R$
larger than the range of their interactions. The wave function has
to satisfy the Efimov boundary condition
\begin{equation}
\psi\;\underset{R\to0}{\propto}\;\frac{1}{R}\sin(s_{0}\log(\Lambda R)),\label{eq:EfimovCondition}
\end{equation}
where $\Lambda$ is the so-called Efimov three-body parameter, which
fixes the accumulated phase of the three-body wave function at low
energy.

The Efimov theory thus relies on only two parameters, $a$ and $\Lambda$,
to fully describe the three-body physics in the low-energy and large
$a$ regime. When normalised in units of $\Lambda$, the trimer energy
spectrum exhibits a universal structure as a function of $a\Lambda$,
represented in Fig.~\ref{fig:Efimov1}. There is an accumulation
point in the spectrum at $a=\infty$, that is when the two-body interaction
is resonant, and the whole spectrum is invariant by a discrete scale
transformation, namely multiplying all distances by $e^{\pi/s_{0}}\approx22.7$,
which follows from Eq.~(\ref{eq:EfimovCondition}). The wave functions
and energies can be calculated numerically by either solving the 3-body
free Schrödinger equation with conditions (\ref{eq:BethePeierls})
and (\ref{eq:EfimovCondition}), or equivalently solving its corresponding
integral equation which is known as the Skorniakov- Ter-Martirosian
equation~\cite{rf:skorniakov},
\begin{equation}
\left(\frac{1}{a}\!-\!\sqrt{{\textstyle \frac{3}{4}}p^{2}\!-\!\varepsilon}\right)\!\! F(p)\!+\!\frac{2}{\pi}\!\!\int_{0}^{P}\!\!\!\!\! dq\ln\!\frac{p^{2}\!+\! q^{2}\!+\! pq\!-\!\varepsilon}{p^{2}\!+\! q^{2}\!-\! pq\!-\!\varepsilon}F(q)=0\label{eq:STM}
\end{equation}
where $F$ is the unknown function (related to the full wave function
$\Psi$), and $\varepsilon=\frac{mE}{\hbar^{2}}$ is the renormalised
energy $E$. Here, the upper bound $P$ of the integral sets the three-body
phase, and can be related to the Efimov three-body parameter by~\cite{rf:naidon}:
\[
P=\Lambda\exp(-\frac{\arctan s_{0}+\pi n}{s_{0}})
\]

\subsection{Non-universal models}

\subsubsection{Non-universal corrections}

The deviations of two-body physics from universality at low energy
are well known and can be encoded in the energy variation of the two-body
phase shift $\delta(E)$, or equivalently a two-body scattering length
$a(E)$. At low energy, we have the following effective-range low-energy
expansion:
\begin{equation}
\frac{1}{a(E)}\equiv-k\cot\delta(E)=\frac{1}{a}-\frac{1}{2}r_{e}k^{2}+\dots\label{eq:EffectiveRange}
\end{equation}
where $k=\sqrt{mE}/\hbar$. The universal limit corresponds to the
first term, which is set by the zero-energy scattering length $a$.
The next order defines the effective range $r_{e}$. It is straightforward
to generalise Eq.~(\ref{eq:STM}) to the non-universal regime by
replacing the zero-energy scattering length $a$ by the energy-dependent
scattering length $a(E)$~\cite{rf:efimov2,rf:naidon}. Likewise,
the cutoff $P$ is expected to be replaced by an energy-dependant
quantity $P(E)$~\cite{rf:naidon}. With these replacements, Eq.~(\ref{eq:STM})
corresponds to the most general contact model with energy-dependent
boundary conditions.

Although the energy dependence of $a(E)$ is generally known, that
of $P(E)$ is presently unknown. It is one of the purposes of this
paper to investigate this dependence from comparison with other models.

\subsubsection{Two-body effective interaction}

Another approach is to replace the real interaction by a simple effective
interaction with the same low-energy spectrum. One possible choice
is a Gaussian potential~\cite{rf:kievsky,rf:Thogersen}:
\begin{equation}
V(r)=-V_{0}e^{-(r/r_{0})^{2}},\label{eq:GaussianPotential}
\end{equation}
which is parametrised by $V_{0}$ and $r_{0}$ to reproduce both the
scattering length and effective range. Another convenient choice is
a separable interaction \cite{rf:lee,rf:jonalasinio,rf:naidon}:
\begin{equation}
\hat{V}=V_{0}\vert\phi\rangle\langle\phi\vert\label{eq:SeparablePotential}
\end{equation}
where the state $\vert\phi\rangle$ can also be chosen to be a Gaussian
function $\phi(r)=e^{-(r/r_{0})^{2}}$ for simplicty. The advantage
of separable potentials is that they have formal similarities with
contact interactions, and lead to a simple integral equation similar
to Eq.~(\ref{eq:STM}):

\begin{eqnarray}
\left(\frac{1}{a(E)}\!-\!\sqrt{{\textstyle \frac{3}{4}}p^{2}\!-\!\varepsilon}\right)\!\! F(p)\!+\quad\quad\qquad\qquad\qquad\qquad\label{eq:STM2}\\
+\!\frac{2}{\pi}\!\!\int_{0}^{\infty}\!\!\!\!\! dq\,\ln\!\frac{G\left(r_{0}^{2}(p^{2}\!+\! q^{2}\!+\! pq\!-\!\varepsilon)\right)}{G\left(r_{0}^{2}(p^{2}\!+\! q^{2}\!-\! pq\!-\!\varepsilon)\right)}e^{\frac{3}{8}r_{0}^{2}(q^{2}-p^{2})}F(q) & = & 0\nonumber 
\end{eqnarray}
where $G(x)=\exp[\int_{x}^{\infty}\frac{e^{-t}}{t}dt]$. One can check
that the integrand of (\ref{eq:STM2}) tends to that of (\ref{eq:STM})
at low momenta, but decays at high momenta $q\gtrsim1/r_{0}$, which
removes the need to introduce an upper bound to the integral.

\section{Realistic and effective calculations for $\,^{4}\mbox{He}_{3}$}

\subsection{Realistic calculations with LM2M2 potentials}

\begin{figure}
\includegraphics{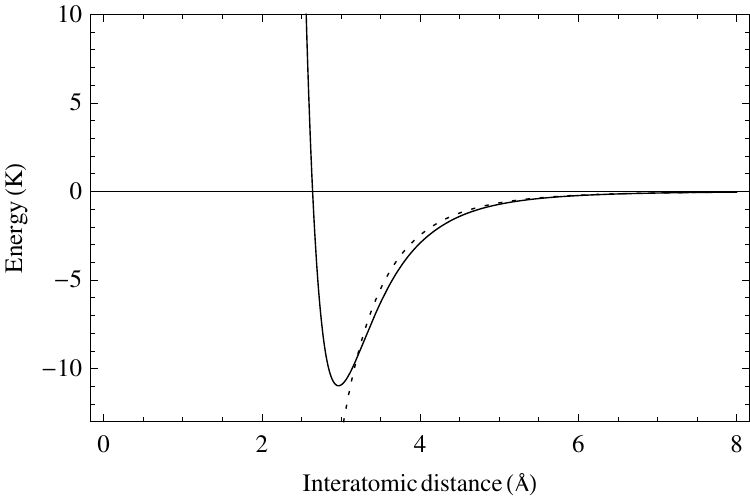}\caption{\label{fig:Potential}LM2M2 potential \cite{rf:aziz} used for the
realistic calculations. The dotted curve indicates the van der Waals
asymptote $-C_{6}/r^{6}$.}
\end{figure}
To model $^{4}\mbox{He}$ interactions realistically, we choose the
LM2M2 potential~\cite{rf:aziz} to describe the two-body interactions.
This potential has a repulsive hard core at short distance, and a
van der Waals tail $-C_{6}/r^{6}$ at large distance, as shown in
Fig.~\ref{fig:Potential}. Its scattering length is 100.01~Å which
is 18.6 times the van der Waals length $l_{\mbox{vdW}}=(mC_{6}/\hbar)^{1/4}$.
The three-body interaction has been shown to bring only small corrections~\cite{rf:Parish,rf:Roeggen,rf:Cencek},
and we neglect it in this study. Thus, in our calculation, the 3-body
phase which fixes the energy of Efimov trimers builds up only from
the LM2M2 two-body interaction.

The 3-body Schrödinger equation with the LM2M2 potential is solved
numerically using the Gaussian expansion method (GEM). This method
was proposed as a means to perform accurate calculation for three-
and four-body systems~\cite{rf:hiyama}. In this method, a well-chosen
set of Gaussian basis function is used, forming an approximately complete
set in a finite coordinate space, so that one can describe accurately
both short-range correlation and the long-range asymptotic behaviour
of the wavefunction for bound systems as well as for scattering states.
It was demonstrated that the GEM provides the same caliber of numerical
precision as, for example, the Faddeev-Yakubovsky method for $^{3}$H
($^{3}$He) and $^{4}$He, and can be used to address various kinds
of few-body problems in atomic, baryonic and quark-level systems~\cite{rf:hiyama}.

\begin{figure}[htb]
\includegraphics[bb=30bp 600bp 480bp 752bp,clip,scale=0.5]{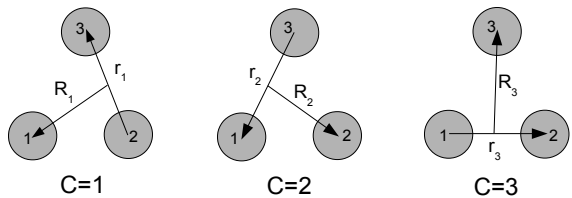}\caption{Jacobi coordinates for all the rearrangement channels ($c=1\sim3$)
of $^{4}$He trimer system. Three $^{4}$He atoms are to be symmetrized.}

\label{fig:jacobi} 
\end{figure}

In order to solve three-body $^{4}$He trimer problem, we use three
sets of Jacobi coordinates illustrated in Fig. \ref{fig:jacobi}.

The Schrödinger equation and the total Hamiltonian are given by 
\begin{equation}
(H-E)\,\Psi_{JM}=0\ ,\label{eq:hamiltonian}
\end{equation}
 
\begin{equation}
H=T+V(r_{1})+V(r_{2})+V(r_{3})\ ,\label{eq:hamiltonian-10-1}
\end{equation}
 where $T$ is the kinetic-energy operator and $V(r_{1})$, $V(r_{2})$
and $V(r_{3})$ are the interactions between two $^{4}$He atoms,
described by the LM2M2 potential $V(r)$.  

The total three-body wavefunction is described as a sum of amplitudes
for all rearrangement channels ($c=1\sim3)$ of Fig. \ref{fig:jacobi}:
\begin{equation}
\Psi_{JM}=\sum_{c=1}^{3}\sum_{n,N}\sum_{\ell,L}C_{nlNL}^{(c)}\big[\phi_{nl}^{(c)}({\bf r}_{c})\psi_{NL}^{(c)}({\bf R}_{c})\big]_{JM}\;.\label{eq:wavefunction}
\end{equation}

We take the functional forms of $\phi_{nlm}({\bf r})$, $\psi_{NLM}({\bf R})$
as 
\begin{eqnarray}
\phi_{nlm}({\bf r}) & = & r^{l}\, e^{-(r/r_{n})^{2}}Y_{lm}({\widehat{{\bf r}}})\;,\nonumber \\
\psi_{NLM}({\bf R}) & = & R^{L}\, e^{-(R/R_{N})^{2}}Y_{LM}({\widehat{{\bf R}}})\;,
\end{eqnarray}
 where the Gaussian range parameters are chosen according to geometrical
progressions: 
\begin{eqnarray}
r_{n} & = & r_{1}a^{n-1}\qquad\enspace(n=1,\dots,n_{{\rm max}})\;,\nonumber \\
R_{N} & = & R_{1}A^{N-1}\quad(N\!=1,\dots,N_{{\rm max}})\;.
\end{eqnarray}
 The eigenenergies $E$ in Eq.\ref{eq:hamiltonian} and the coefficients
$C$ in Eq.\ref{eq:wavefunction} are determined by the Rayleigh-Ritz
variational method.

\begin{figure}
\includegraphics{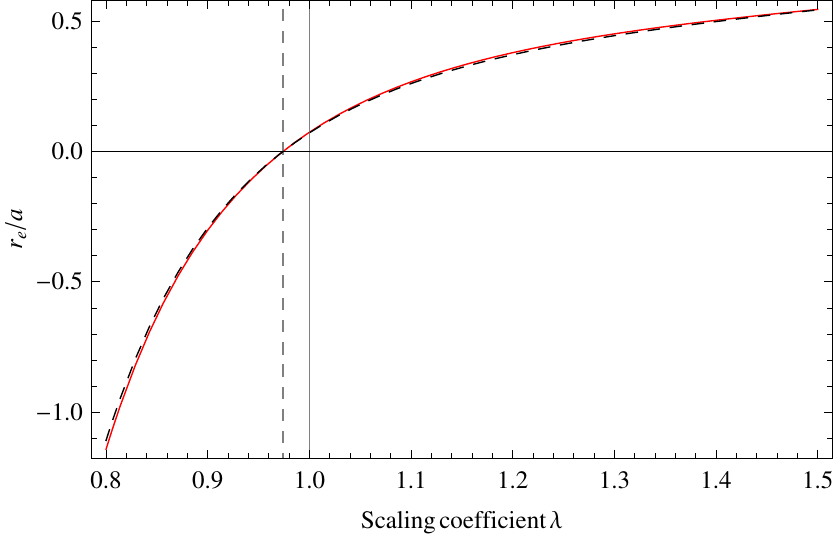}

\caption{\label{fig:scatteringLength}Ratio between the effective range $r_{e}$
and scattering length $a$ of the scaled LM2M2 potential, as a function
of the scaling coefficient $\lambda$. The vertical solid line indicates
the physical value ($\lambda=1$). The dashed vertical line indicates
the scaling where the scattering length diverges ($1/a\to0$). The
dashed curve shows the result based on the analytical formula (\ref{eq:BoGaoFormula})
using the values of $\bar{a}$ and $a$ as a function of $\lambda$.}
\end{figure}

Although the scattering length of $^{4}\mbox{He}$ is already large
compared to the range of its two-body potential $V(r)$, we did several
calculations for rescaled potentials $\lambda V(r)$ in order to cause
a divergence of the scattering length, as was done in previous studies~\cite{rf:Cornelius,rf:esrylin,rf:Gonzalez.}.
This enables us to mimic the broad Feshbach resonances used in ultra-cold
atom experiments~\cite{rf:chin}, and better appreciate the Efimov
structure of the spectrum. The scattering length $a$ diverges for
$\lambda=0.97412$, while the physical results for real $^{4}\mbox{He}$
correspond to $\lambda=1$. As $\lambda$ is varied near the divergence
of $a$, the effective range $r_{e}$ also changes but always remains
positive and on the order of the scaled van der Waals length $\lambda^{1/4}l_{\mbox{vdW}}$,
and the ratio $r_{e}/a$ remains a monotonic function of $\lambda$,
as shown in Fig.~\ref{fig:scatteringLength}. For this reason, we
choose to report our results as a function of the scattering length
in units of the effective range, rather than $\lambda$ itself.

\begin{figure}
\includegraphics{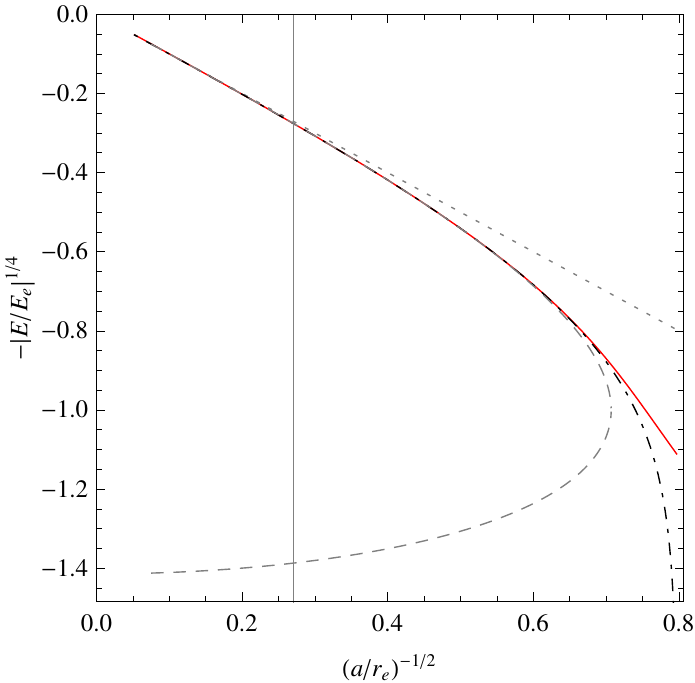}

\caption{\label{fig:dimer}Helium 4 dimer energy $E$ as a function of scattering
length $a$. To clarify the figure, these quantities are normalised
to $E_{e}=\hbar^{2}/mr_{e}^{2}$ and $r_{e}$, and raised to the power
$1/4$ and $-1/2$, respectively. The solid red curve corresponds
to the dimer of the LM2M2 potential scaled by a coefficient $\lambda$.
The physical scattering length of helium 4 ($\lambda=1$) is indicated
by the vertical gray line. The dotted line represents the universal
limit of small binding energy and large scattering length {[}Eq.~(\ref{eq:TwoBodyEnergy})
with $a(E)\to a${]}. The dashed curve represents the two solutions
of the effective range approximation of Eq.~(\ref{eq:TwoBodyEnergy}).
The dotted-dashed curve corresponds to the results of the separable
Gaussian potential (\ref{eq:SeparablePotential}).}
\end{figure}

\begin{figure}
\includegraphics{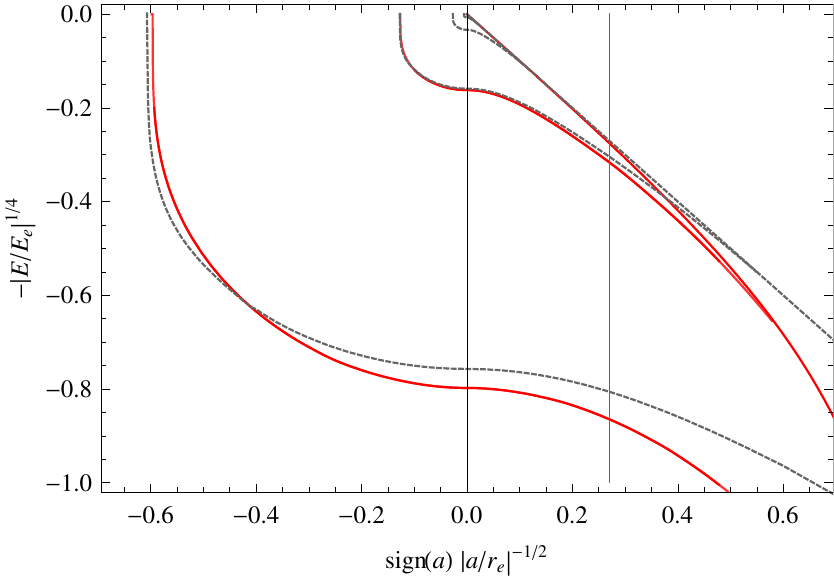}

\caption{\label{fig:Efimov1}Efimov spectrum for helium 4: trimer energy $E$
as a function of the scattering length $a$. To clarify the figure,
these quantities are normalised to $E_{e}=\hbar^{2}/mr_{e}^{2}$ and
$r_{e}$, and raised to the power $1/4$ and $-1/2$, respectively.
The zero of the energy axis corresponds to the three-body threshold.
The red solid curves are the energy curves obtained by scaling the
LM2M2 potential. Both trimers and dimer (rightmost curve) are displayed.
The vertical gray line indicates the scattering length corresponding
to the unscaled potential, \emph{i.e.} physical helium 4. The dashed
curves correspond to the Efimov trimer spectrum according to the universal
theory. The 3-body parameter is adjusted to match the first excited
trimer state. The straight dashed line corresponds to the dimer energy
in the universal limit of large scattering length.}
\end{figure}

The LM2M2 potential supports one two-body bound state and its energy
variation with the scattering length is represented in Fig.~\ref{fig:dimer}.
When the scattering length becomes smaller than its physical value
($\lambda\gtrsim1.0$), this dimer energy significantly deviates from
the universal limit of small binding energy and large scattering length
(Eq.~(\ref{eq:TwoBodyEnergy}) with $a(E)\to a$). The results for
three atoms are shown in Fig.~\ref{fig:Efimov1}. Our results are
in very good agreement with the most accurate calculations using the
LM2M2 potential~\cite{rf:BlumeGreene,rf:Carbonell}. One can see
that the first two trimers' energies qualitatively follow Efimov's
universal spectrum, but as in the two-body case, there are significant
deviations for small scattering lengths and deep energies.


\subsection{Calculation with non-universal corrections}

We first attempt to reproduce the previous results by including non-universal
corrections to the Efimov theory. For each value of $\lambda$, we
can determine the energy dependence $a(E)$ of the scattering length
of the scaled LM2M2 potential $\lambda V(r)$. The energy $E_{2B}$
of the two-body bound state is then readily obtained from the equation:
\begin{equation}
E_{2B}=-\frac{\hbar^{2}}{m[a(E_{2B})]^{2}}\label{eq:TwoBodyEnergy}
\end{equation}
The low-energy expansion of $a(E)$ up to the effective-range term
(right-hand side of Eq. (\ref{eq:EffectiveRange})) can already reproduce
the realistic two-body energy for $a\gtrsim2.5r_{e}$- see Fig.~\ref{fig:dimer}.
However at this order of expansion there are actually two solutions
to Eq.~(\ref{eq:TwoBodyEnergy}), the lowest-energy solution being
unphysical. The two solutions merge and disappear at $a=2r_{e}$.
The presence of this extra dimer is an artefact which completely changes
the 3-body physics. Note that this problem does not occur for negative
effective ranges, as in the case of narrow Feshbach resonances~\cite{rf:Petrov}.
To remedy this problem, we consider an improved analytical expression
for $a(E)$ that is obtained from the separable potential Eq.~(\ref{eq:SeparablePotential})
adjusted to reproduce the scattering length and effective range of
the scaled LM2M2 potential. Then there is only one solution to Eq.~(\ref{eq:TwoBodyEnergy})
and its energy matches very well that of the LM2M2 potential - see
Fig.~\ref{fig:dimer}. Interestingly, the agreement is even slightly
better than the effective-range approximation itself.

\begin{figure}
\includegraphics{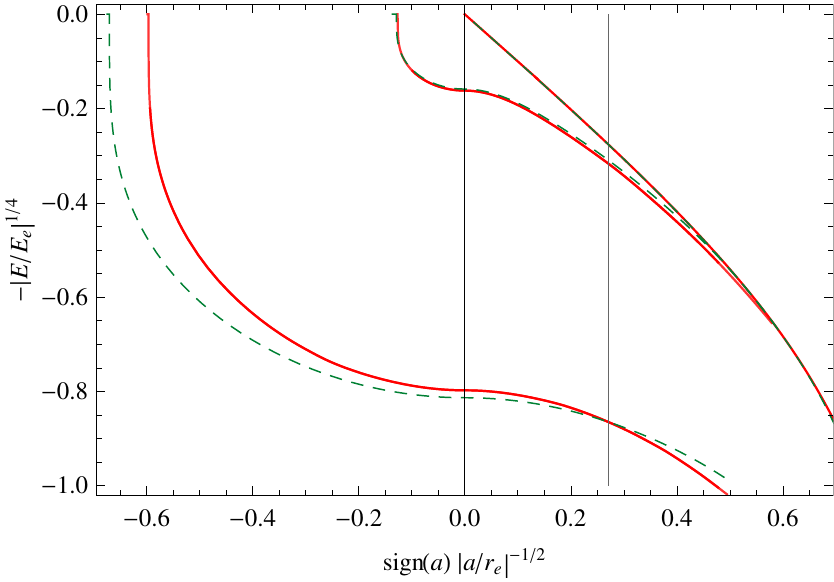}

\caption{\label{fig:Efimov2}Similar plot to Fig.~\ref{fig:Efimov1}. Here
again, the helium 4 curves are indicated by solid red curves, and
results taking into account two-body corrections to the universal
theory are indicated by dashed curves. }
\end{figure}

\begin{figure}
\includegraphics{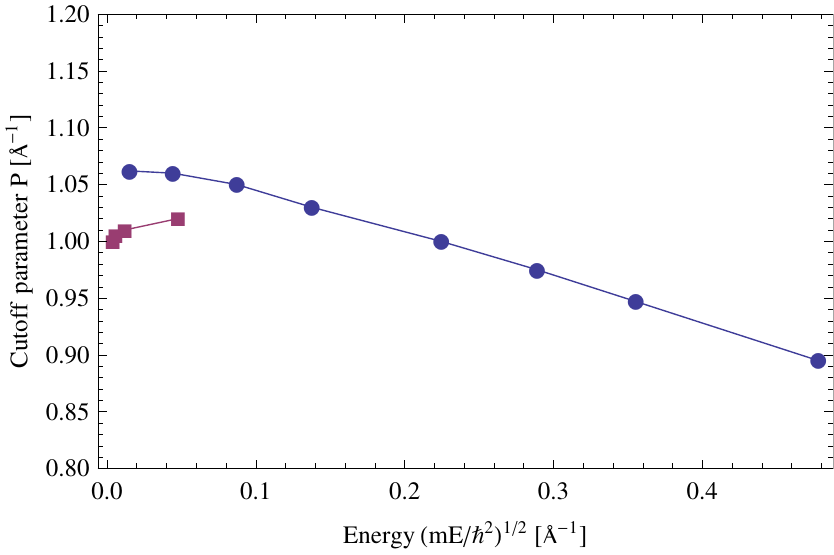}

\caption{\label{fig:ThreeBodyParameter}Three-body cutoff parameter $P$ adjusted
to get agreement with the LM2M2 results, as a function of energy,
for the ground-state trimer (dots) and the excited-state trimer (squares).
One can see that their variations are inconsistent.}
\end{figure}

Having set $a(E)$ to properly describe two-body physics, we perform
trimer calculations using Eq.~(\ref{eq:STM}) with a fixed cutoff
$P$ whose value is adjusted to reproduce the second 3-body dissociation
point - the point where the first excited trimer reaches the threshold.
One can see in Fig.~\ref{fig:Efimov2} that the corrections bring
some improvement, but the agreement with the LM2M2 results is only
partial. We then determined the required variation of the cutoff $P$
in order to obtain perfect agreement with the LM2M2 results for both
the ground and first excited trimer states. The variation is represented
in Fig.~\ref{fig:ThreeBodyParameter} and is inconsistent for both
states. To remove the inconsistency, we can assume more generally
that $P$ depends on both the energy and the scattering length, but
no clear pattern arises from such considerations. It should also be
noted that the required variation of $P$ is dependent on the choice
of the high-energy behaviour of $a(E)$, in other words it is model-dependent.
Thus, while it can be a practical way to characterize non-universal
observations, as done in Refs.~\cite{rf:nakajima,rf:nakajima2,rf:naidon},
it does not seem to be very meaningful.

\subsection{Calculations with a separable potential}

We then attempt to reproduce the $^{4}\mbox{He}$ results with a simple
two-body potential $V$ having the same low-energy properties as the
LM2M2 potential. For both the Gaussian potential (\ref{eq:GaussianPotential})
and separable potential (\ref{eq:SeparablePotential}), we proceed
as follows.

\begin{figure}
\includegraphics{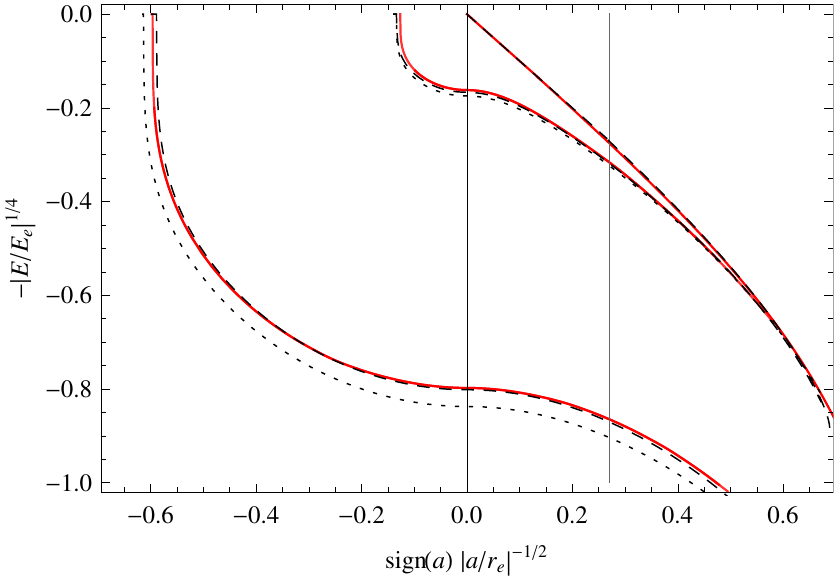}

\caption{\label{fig:Efimov3}Similar plot to Fig.~\ref{fig:Efimov1} for the
calculations with the gaussian potential Eq.~(\ref{eq:GaussianPotential}).
The dotted curves correspond to two-body interactions only, while
the dashed curves correspond to calculation with an additional three-body
interaction.}
\end{figure}

\begin{figure}
\includegraphics{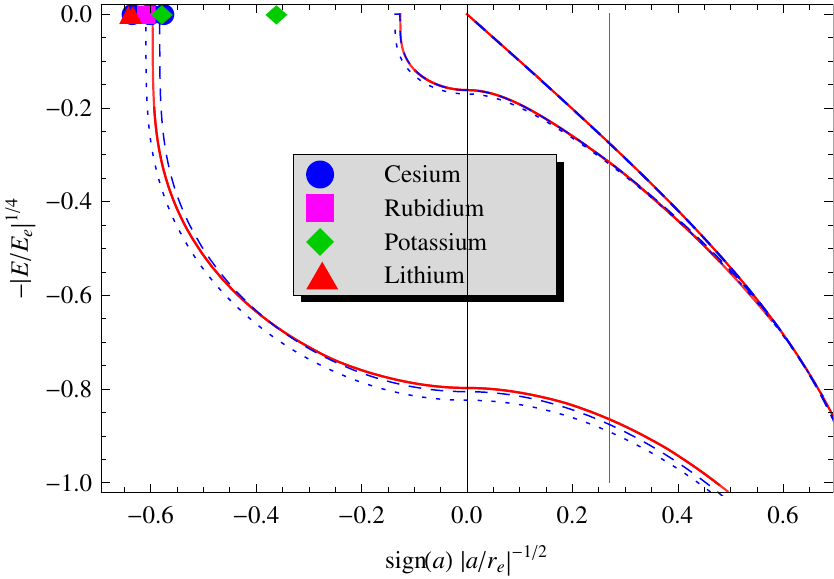}

\caption{\label{fig:Efimov4}Similar plot to Fig.~\ref{fig:Efimov1} for the
calculations with the separable gaussian potential Eq.~(\ref{eq:SeparablePotential}).
The dotted curves correspond to two-body interactions only, while
the dashed curves correspond to calculation with an additional three-body
interaction. The symbols show the measured dissociation point of the
ground-state trimer for different species.}
\end{figure}

For each $\lambda$, we adjust the parameters of the potential so
that the scattering length and effective range coincide with those
of the scaled LM2M2 potential. As mentioned before, the binding energy
of the two-body bound state then matches very well that of the LM2M2
bound state over a wide range of energies - see Fig.~\ref{fig:dimer}.

We then calculate the three-body bound states with the adjusted potential.
Remarkably, we already find relatively good agreement with the LM2M2
three-body states - see Figs.~\ref{fig:Efimov3} and \ref{fig:Efimov4}.
The values differ essentially by an energy shift, which is not unexpected
since the short-range three-body phase may not be properly set by
these simple two-body interactions.

We finally add a three-body interaction (also chosen to be Gaussian
or separable), and adjust its strength in order to reproduce the second
three-body dissociation point. We then find very good overall agreement
with the LM2M2 calculations, as shown in Figs.~\ref{fig:Efimov3}
and~\ref{fig:Efimov4}.

\subsection{Role of the effective range}

The previous results clearly indicate that the non-universal variations
of the trimer energies can be reproduced to a great extent from the
knowledge of the effective range. The role of the effective range
was pointed out before in a previous studies~\cite{rf:efimov2,rf:Thogersen,rf:platter,rf:Ji}.
In particular, Ref.~\cite{rf:Thogersen} also looked at deviations
from universality with finite-range potentials. This study considered
the scattering length $a_{\mbox{diss}}$ for which the trimer dissociates.
The relative variation of this quantity with respect to its value
$a_{\mbox{diss}}^{\prime}$ in the universal spectrum was found to
be:
\begin{equation}
\frac{a_{\mbox{diss}}-a_{\mbox{diss}}^{\prime}}{a_{\mbox{diss}}^{\prime}}=C\left(\frac{r_{e}}{a}\right)_{\mbox{diss}}\label{eq:ThogersonFormula}
\end{equation}
with $C=1.3\pm0.4$. However, we could not completely confirm this
formula. While we found the value $C\approx0.99$ with the Gaussian
potential, which is consistent with the results presented in Ref.~\cite{rf:Thogersen}
for the same Gaussian potential, the present calculation with the
LM2M2 potential gives $C\approx0.58$. The value therefore seems to
be somewhat dependent on the type of potential.

It is interesting to note that the value of the dissociation point
$a_{\mbox{diss}}$ itself, rather than its non-universal variation,
was found experimentally to be in a narrow range. Experimentalists
have measured the ground-state trimer dissociation point near several
divergences of the scattering lengths~\cite{rf:Berninger}, and for
different species~\cite{rf:kraemer,rf:pollack,rf:gross,rf:gross2,rf:zaccanti,rf:ferlaino,rf:Berninger},
and found that it almost always occurs around $a_{\mbox{diss}}\approx-9.9\bar{a}$,
where $\bar{a}=2\pi/\Gamma(1/4)^{2}l_{\mbox{vdW}}$ is the average
scattering length of van der Waals potentials~\cite{rf:chin}. Since
there is an approximate relation between $\bar{a}$ and the effective
range $r_{e}$~\cite{rf:Gao,rf:Flambaum}:
\begin{equation}
\frac{r_{e}}{a}=\frac{2}{3}\frac{\Gamma(1/4)^{4}}{(2\pi)^{2}}\frac{\bar{a}}{a}\left(\left(\frac{\bar{a}}{a}\right)^{2}+\left(\frac{\bar{a}}{a}-1\right)^{2}\right),\label{eq:BoGaoFormula}
\end{equation}
this corresponds to $a_{\mbox{diss}}/r_{e}\approx-2.76$. Experimental
measurements of $a_{\mbox{diss}}/r_{e}$ for different species are
represented in Fig.~\ref{fig:Efimov4}. It is quite remarkable that
the scaled helium potential also gives a dissociation point $a_{\mbox{diss}}/r_{e}\approx-2.8$2
(or equivalently $a_{\mbox{diss}}/\bar{a}=-10.26$) which lies in
the same narrow range. Furthermore, the model potentials considered
in this paper (Gaussian and separable potential), without any 3-body
interaction, also give a consistent value $a_{\mbox{diss}}/r_{e}\approx-2.70$
of the dissociation point. This shows that the use of these model
potentials to interpret experiments leads to a good estimate of the
dissociation point, as noted before~\cite{rf:jonalasinio,rf:naidon}.
Why it does so is however a puzzling question, since we know from
calculations with other or deeper potentials, that the dissociation
point can change significantly~\cite{rf:dincao}.This suggests that,
under some condition yet to be discovered, the effective range may
not only determine the non-universal variations, but also the three-body
parameter, thereby determining the full three-body spectrum.

\section{Conclusion}

In this work, we have studied the Efimov physics of three helium-4
atoms, as a simple but realistic example to understand the non-universal
variations of the trimer energy with respect to the scattering length.
We found that non-universal two-body corrections to the universal
theory alone are not sufficient to fully reproduce these variations,
and a variable three-body parameter is needed. However the variations
of this three-body parameter do not seem to follow any simple rule,
and are model-dependent. On the other hand, \emph{ad hoc} but simple
two-body potentials such as separable Gaussian potentials adjusted
to have the same scattering length and effective range reproduce remarkably
well the trimer energy non-universal variations, and can constitute
relatively accurate substitutes for single-channel realistic interactions
provided that a small and localised three-body force is introduced
to properly shift the energy.  This indicates that the non-universal
deviations which were observed in \cite{rf:nakajima,rf:nakajima2}
and could not be explained by these effective potential models are
likely to be due to more subtle effects such as multichannel coupling,
non-trivial effects of three-body forces, unless they simply result
from underestimated uncertainties either in the scattering lengths
or experimental data. 

Provided that such subtle effects are absent, our results also suggest
that in general, beyond the usual Efimov universal scenario occurring
for higher excited trimers, the whole spectrum follows a more specific
class of universality determined only by the scattering length, the
effective range, and the strength of a three-body localised force
setting the three-body parameter, i.e. the position of highly-excited
trimers.

We thank F.~Ferlaino, R.~Grimm, P.S.~Julienne, J.~D'Incao, C.H.~Greene
and E.~Braaten for helpful discussions. Some of the numerical calculations
were performed on the HITACHI SR11000 at KEK.


\begin{thebibliography}{References}
\bibitem{rf:efimov}V.~N.~Efimov, Sov. J. Nucl. Phys. \textbf{12},
589 (1970) ; V.~Efimov, Nucl. Phys. A, 210 , 157 (1973).

\bibitem{rf:ferlaino}F.~Ferlaino and R.~Grimm, Physics \textbf{3},
9 (2010).

\bibitem{rf:kraemer}T. Kraemer, et al. Nature \textbf{440}, 315\textendash{}318
(2006).

\bibitem{rf:ottenstein}T.~B.~Ottenstein, et al., Phys. Rev. Lett.
\textbf{101}, 203202 (2008).

\bibitem{rf:williams}J.~R.~Williams, et al., Phys. Rev. Lett.
\textbf{103}, 130404 (2009).

\bibitem{rf:zaccanti}M.~Zaccanti, et al., Nature Physics \textbf{5},
586 - 591 (2009).

\bibitem{rf:barontini}G.~Barontini et al., Phys. Rev. Lett. \textbf{103},
043201 (2009).

\bibitem{rf:wenz}A.~N.~Wenz, et al., Phys. Rev. \textbf{A} 80,
040702 (2009).

\bibitem{rf:pollack}S.~E. Pollack, D.~Dries, R.~G. Hulet, Science
\textbf{326}, 1683 (2009).

\bibitem{rf:gross}N.~Gross, Z.~Shotan, S.~Kokkelmans, and L.~Khaykovich,
Phys. Rev. Lett. \textbf{103}, 163202 (2009).

\bibitem{rf:huckans} J.~H.~Huckans et al., Phys. Rev. Lett. \textbf{102},
165302 (2009).

\bibitem{rf:lompe1}T.~Lompe, et al., Phys. Rev. Lett. \textbf{105},
103201 (2010).

\bibitem{rf:lompe2}T.~Lompe, et al., Science \textbf{330}, 940
(2010).

\bibitem{rf:gross2}N.~Gross, Z.~Shotan, S.~Kokkelmans, and L.~Khaykovich,
Phys. Rev. Lett. \textbf{105}, 103203 (2010).

\bibitem{rf:nakajima}S.~Nakajima, M.~Horikoshi, T.~Mukaiyama,
P.~Naidon, and M.~Ueda, Phys. Rev. Lett. \textbf{105}, 023201 (2010).

\bibitem{rf:nakajima2}S.~Nakajima, M.~Horikoshi, T.~Mukaiyama,
P.~Naidon, and M.~Ueda, Phys. Rev. Lett. \textbf{106}, 143201 (2011).

\bibitem{rf:Berninger}M. Berninger et al., Phys. Rev. Lett. \textbf{107},
120401 (2011).

\bibitem{rf:platter}Lucas Platter, Chen Ji, and Daniel R. Phillips,
Phys. Rev. A \textbf{79}, 022702 (2009).

\bibitem{rf:Ji}C. Ji, L. Platter, and D. R. Phillips, Europhys. Lett.
\textbf{92}, 13003 (2010).

\bibitem{rf:lee}M.~D.~Lee, T.~Köhler, and P.~S.~Julienne, Phys.
Rev. A \textbf{76}, 012720 (2007).

\bibitem{rf:jonalasinio}M.~Jona-Lasinio and L.~Pricoupenko, Phys.
Rev. Lett. \textbf{104}, 023201 (2010).

\bibitem{rf:naidon}P. Naidon and M. Ueda, Comptes Rendus Physique,
\textbf{12}, iss.~1, p.~13-26 (2011) {[}arXiv:1008.2260 (2010){]}.

\bibitem{rf:dincao}J.~P.~D\textquoteright{}Incao, C.~H.~Greene,
and B.~D.~Esry, J. Phys. B \textbf{42}, 044016 (2009).

\bibitem{rf:Cornelius}T. Cornelius and W. Glöckle, J. Chem. Phys.
\textbf{85}, 3906 (1986).

\bibitem{rf:esrylin}B. D. Esry, C. D. Lin, and Chris H. Greene, Phys.
Rev. A \textbf{54}, 394 (1996).

\bibitem{rf:NielsenHelium}E. Nielsen, D.V. Fedorov, and A.S. Jensen,
J. Phys. B \textbf{31}, 4085 (1998).

\bibitem{rf:Gonzalez.}T. González-Lezana et al, Phys. Rev. Lett.
\textbf{82}, 1648 (1999).

\bibitem{rf:BraatenHelium}Eric Braaten, H.-W. Hammer, Phys. Rev.
A \textbf{67,} 042706 (2003).

\bibitem{rf:skorniakov}G. V. Skorniakov and K. A. Ter-Martirosian,
Sov. Phys. JETP \textbf{4}, 648 (1957).

\bibitem{rf:efimov2}V.~Efimov, Phys. Rev. C \textbf{44}, 2303 (1991).

\bibitem{rf:kievsky}A. Kievsky, E. Garrido, C. Romero-Redondo, and
P. Barletta, Few-Body Syst DOI 10.1007/s00601-011-0226-9 (2011).

\bibitem{rf:Thogersen}M. Thøgersen, D. V. Fedorov, and A. S. Jensen,
Phys. Rev. A \textbf{78}, 020501R (2008).

\bibitem{rf:aziz}R.A.~Aziz, M.J.~Slaman, J. Chem. Phys. \textbf{94},
8047 (1991).

\bibitem{rf:Parish}C.A. Parish and C.E. Dykstra, J. Chem. Phys. \textbf{101},
7618 (1994).

\bibitem{rf:Roeggen}I. Røeggen and J. Almlöf, J. Chem. Phys. \textbf{102},
7095 (1995).

\bibitem{rf:Cencek}W. Cencek, M. Jeziorska, O. Akin-Ojo, and K. Szalewicz,
J. Phys. Chem. A, \textbf{111} (44), 11311 (2007).

\bibitem{rf:hiyama}E. Hiyama, Y. Kino, and M. Kamimura, Prog. Part.
and Nucl. Phys. \textbf{51}, 223 (2003).

\bibitem{rf:chin}C.~Chin, R.~Grimm, P.~Julienne, and E.~Tiesinga,
Rev. Mod. Phys. \textbf{82}, 1225 (2010).

\bibitem{rf:BlumeGreene}D. Blume and C. H. Greene, J. Chem. Phys.
\textbf{112}, 8053 (2000).

\bibitem{rf:Carbonell}R.~Lazauskas, J.~Carbonell, Phys. Rev. A
\textbf{73}, 062717 (2006).

\bibitem{rf:Petrov}D.~S.~Petrov, Phys. Rev. Lett. \textbf{93},
143201 (2004). 

\bibitem{rf:Gao}B. Gao, Phys. Rev. A \textbf{58,} 4222 (1998).

\bibitem{rf:Flambaum}V.V. Flambaum, G.F. Gribakin, and C. Harabati,
Phys. Rev. A \textbf{59,} 1998 (1999).\end{thebibliography}
\end{document}